\include{epsf}
\documentstyle[times,dl]{article}

\begin{document}
\title{Content-Based Book Recommending \\ Using Learning for Text Categorization}
\author{ Raymond J. Mooney  \\ 
         Department of Computer Sciences 
	 \smallskip \\
         Loriene Roy \\
         Graduate School of Library and Information Science 
         \smallskip \\
         University of Texas \\
         Austin, TX 78712
         \smallskip \\
         Email : mooney@cs.utexas.edu, loriene@gslis.utexas.edu
        }

\maketitle 
\abstract
Recommender systems improve access to relevant products and information by
making personalized suggestions based on previous examples of a user's likes and
dislikes.  Most existing recommender systems use social filtering methods that
base recommendations on other users' preferences. By contrast, content-based
methods use information about an item itself to make suggestions.  This approach
has the advantage of being able to recommended previously unrated items to
users with unique interests and to provide explanations for its
recommendations. We describe a content-based book recommending system that
utilizes information extraction and a machine-learning algorithm for text
categorization.  Initial experimental results demonstrate that this approach
can produce accurate recommendations.


\section{INTRODUCTION}

There is a growing interest in {\it recommender systems} that suggest music,
films, books, and other products and services to users based on examples of
their likes and dislikes \cite{maes:cacm94,resnik:cacm97,rec-wkshp98}.  A
number of successful startup companies like Firefly, Net Perceptions, and
LikeMinds have formed to provide recommending technology.  On-line book stores
like Amazon and BarnesAndNoble have popular recommendation services, and many
libraries have a long history of providing {\it reader's advisory} services
\cite{baker:colbuild93,mccook:book93}.  Such services are important since
readers' preferences are often complex and not readily reduced to keywords or
standard subject categories, but rather best illustrated by example.  Digital
libraries should be able to build on this tradition of assisting readers by
providing cost-effective, informed, and personalized automated recommendations
for their patrons.

Existing recommender systems almost exclusively utilize a form of computerized
matchmaking called {\it collaborative} or {\it social filtering}.  The system
maintains a database of the preferences of individual users, finds other users
whose known preferences correlate significantly with a given patron, and
recommends to a person other items enjoyed by their matched patrons.  This
approach assumes that a given user's tastes are generally the same as another
user of the system and that a sufficient number of user ratings are available.
Items that have not been rated by a sufficient number of users cannot be
effectively recommended.  Unfortunately, statistics on library use indicate
that most books are utilized by very few patrons \cite{kent:book79}.
Therefore, collaborative approaches naturally tend to recommend popular titles,
perpetuating homogeneity in reading choices. Also, since significant
information about other users is required to make recommendations, this
approach raises concerns about privacy and access to proprietary customer data.

Learning individualized profiles from descriptions of examples ({\it
content-based recommending} \cite{balabanovic:cacm97}), on the other hand,
allows a system to uniquely characterize each patron without having to match
their interests to someone else's.  Items are recommended based on information
about the item itself rather than on the preferences of other users.  This also
allows for the possibility of providing explanations that list content features
that caused an item to be recommended; potentially giving readers confidence in
the system's recommendations and insight into their own preferences.  Finally,
a content-based approach can allow users to provide initial subject information
to aid the system.

Machine learning for text-categorization has been applied to content-based
recommending of web pages \cite{pazzani:aaai96} and newsgroup messages
\cite{lang:ml95}; however, to our knowledge has not previously been applied to
book recommending.  We have been exploring content-based book recommending by
applying automated text-categorization methods to semi-structured text
extracted from the web.  Our current prototype system, {\sc Libra} (Learning
Intelligent Book Recommending Agent), uses a database of book information
extracted from web pages at Amazon.com.  Users provide 1--10 ratings for a
selected set of training books; the system then learns a profile of the user
using a  Bayesian learning algorithm and produces a ranked list of the
most recommended additional titles from the system's catalog.

As evidence for the promise of this approach, we present initial experimental
results on several data sets of books randomly selected from particular genres
such as mystery, science, literary fiction, and science fiction and rated by
different users.  We use standard experimental methodology from machine
learning and present results for several evaluation metrics on independent test
data including rank correlation coefficient and average rating of top-ranked
books.

The remainder of the paper is organized as follows.  Section 2 provides an
overview of the system including the algorithm used to learn user profiles.
Section 3 presents results of our initial experimental evaluation of the
system.  Section 4 discusses topics for further research, and section 5
presents our conclusions on the advantages and promise of content-based book
recommending.

\section{SYSTEM DESCRIPTION}

\subsection{Extracting Information and Building a Database}

First, an Amazon subject search is performed to obtain a list of
book-description URL's of broadly relevant titles.  {\sc Libra} then downloads
each of these pages and uses a simple pattern-based information-extraction
system to extract data about each title.  Information extraction (IE) is the
task of locating specific pieces of information from a document, thereby
obtaining useful structured data from unstructured text
\cite{lehnert:aimag91,muc:muc95}.  Specifically, it involves finding a set of
substrings from the document, called {\it fillers}, for each of a set of
specified {\it slots}.  When applied to web pages instead of natural language
text, such an extractor is sometimes called a {\it wrapper}
\cite{kushmerick:ijcai97}.  The current slots utilized by the recommender are:
title, authors, synopses, published reviews, customer comments, related
authors, related titles, and subject terms. Amazon produces the information
about related authors and titles using collaborative methods; however, {\sc
Libra} simply treats them as additional content about the book.  Only books
that have at least one synopsis, review or customer comment are retained as
having adequate content information. A number of other slots are also extracted
(e.g. publisher, date, ISBN, price, etc.) but are currently not used by the
recommender.  We have initially assembled databases for literary fiction (3,061
titles), science fiction (3,813 titles), mystery (7,285 titles), and science
(6,177 titles).

Since the layout of Amazon's automatically generated pages is quite regular, a
fairly simple extraction system is sufficient.  {\sc Libra}'s extractor employs
a simple pattern matcher that uses pre-filler, filler, and post-filler patterns
for each slot, as described by \cite{mecaliff:ss98}.  In other applications,
more sophisticated information extraction methods and inductive learning of
extraction rules might be useful \cite{cardie:aimag97}.  

The text in each slot is then processed into an unordered bag of words (tokens)
and the examples represented as a vector of bags of words (one bag for each
slot).  A book's title and authors are also added to its own related-title and
related-author slots, since a book is obviously ``related'' to itself, and this
allows overlap in these slots with books listed as related to it.  Some minor
additions include the removal of a small list of stop-words, the preprocessing
of author names into unique tokens of the form first-initial\_last-name and the
grouping of the words associated with synopses, published reviews, and customer
comments all into one bag (called ``words'').

\subsection{Learning a Profile}

Next, the user selects and rates a set of training books.  By searching for
particular authors or titles, the user can avoid scanning the entire database
or picking selections at random.  The user is asked to provide a discrete 1--10
rating for each selected title.  

The inductive learner currently employed by {\sc Libra} is a bag-of-words naive
Bayesian text classifier \cite{mitchell:book97} extended to handle a vector of
bags rather than a single bag.  Recent experimental results
\cite{joachims:ml97,mccallum:textcat-wkshp98} indicate that this relatively
simple approach to text categorization performs as well or better than many
competing methods.  {\sc Libra} does not attempt to predict the exact numerical
rating of a title, but rather just a total ordering (ranking) of titles in
order of preference.  This task is then recast as a probabilistic binary
categorization problem of predicting the probability that a book would be rated
as positive rather than negative, where a user rating of 1--5 is interpreted
as negative and 6--10 as positive.  As described below, the exact numerical
ratings of the training examples are used to weight the training examples when
estimating the parameters of the model.

Specifically, we employ a multinomial text model
\cite{mccallum:textcat-wkshp98}, in which a document is modeled as an ordered
sequence of word events drawn from the same vocabulary, $V$.  The ``naive
Bayes'' assumption states that the probability of each word event is dependent
on the document class but independent of the word's context and position.  For
each class, $c_j$, and word or token, $w_k \in V$, the
probabilities, $P(c_j)$ and $P(w_k | c_j)$ must be estimated from the training
data.  Then the posterior probability of each class given a document, $D$, is
computed using Bayes rule:
\begin{equation}
P(c_j | D) = \frac{P(c_j)}{P(D)} \prod_{i=1}^{|D|} P(a_i | c_j) 
\end{equation}
where $a_i$ is the $i$th word in the document, and $|D|$ is the length of the
document in words. Since for any given document, the prior $P(D)$ is a
constant, this factor can be ignored if all that is desired is a ranking rather
than a probability estimate.  A ranking is produced by sorting documents by their
odds ratio, $P(c_1 | D)/P(c_0 | D)$, where $c_1$ represents the
positive class and $c_0$ represents the negative class.  An example is
classified as positive if the odds are greater than 1, and negative otherwise.

In our case, since books are represented as a vector of ``documents,'' $d_m$,
one for each slot (where $s_m$ denotes the $m$th slot), the probability of each
word given the category and the slot, $P(w_k | c_j, s_m)$, must be estimated
and the posterior category probabilities for a book, $B$, computed using:
\begin{equation}
P(c_j | B) = \frac{P(c_j)}{P(B)} \prod_{m=1}^{S}\prod_{i=1}^{|d_m|} P(a_{mi} | c_j, s_m)
\end{equation}
where $S$ is the number of slots and $a_{mi}$ is the $i$th word in the $m$th slot.

Parameters are estimated from the training examples as follows.  Each of the
$N$ training books, $B_e$ ($1 \leq e \leq N$) is given two real weights, $0
\leq \alpha_{ej} \leq 1$, based on scaling it's user rating, $r$ ($1 \leq r \leq
10$) : a positive weight, $\alpha_{e1} = (r - 1)/9$, and a negative weight
$\alpha_{e0} = 1 - \alpha_{e1}$.  If a word appears $n$ times in an example
$B_e$, it is counted as occurring $\alpha_{e1} n$ times in a positive example
and $\alpha_{e0} n$ times in a negative example.  The model parameters are
therefore estimated as follows:
\begin{equation}
P(c_j) = \sum_{e=1}^{N}\alpha_{ej} / N
\end{equation}
\begin{equation}
P(w_k | c_j, s_m) = \sum_{e=1}^{N} \alpha_{ej} n_{kem} / L(c_j, s_m)
\end{equation}
where $n_{kem}$ is the count of the number of times word $w_k$ appears
in example $B_e$ in slot $s_m$, and
\begin{equation}
L(c_j, s_m) = \sum_{e=1}^{N}\alpha_{ej}|d_m|
\end{equation}
denotes the total weighted length of the documents in category $c_j$ and
slot $s_m$.

These parameters are ``smoothed'' using Laplace estimates to avoid zero
probability estimates for words that do not appear in the limited training
sample by redistributing some of the probability mass to these items using the
method recommended in \cite{kohavi:ecml97}.  Finally, calculation with
logarithms of probabilities is used to avoid underflow.  

The computational complexity of the resulting training (testing) algorithm is
linear in the size of the training (testing) data. Empirically, the system is
quite efficient. In the experiments on the {\sc Lit1} data described below, the
current Lisp implementation running on a Sun Ultra 1 trained on 20 examples in
an average of 0.4 seconds and on 840 examples in an average of 11.5 seconds,
and probabilistically categorized new test examples at an average rate of about
200 books per second.  An optimized implementation could no doubt significantly
improve performance even further.

A profile can be partially illustrated by listing the features most indicative
of a positive or negative rating.  Table \ref{pos-profile} presents the top 20
features for a sample profile learned for recommending science books.  {\it
Strength} measures how much more likely a word in a slot is to appear in a
positively rated book than a negatively rated one, computed as:
\begin{equation}
Strength(w_k, s_j) = log(P(w_k | c_1,s_j)/P(w_k | c_0,s_j))
\end{equation}

\begin{table}
\begin{footnotesize}
\begin{center}
\begin{tabular}{llr}
Slot & Word & Strength \\
\hline
WORDS & ZUBRIN               &   9.85 \\
WORDS & SMOLIN               &   9.39 \\
WORDS & TREFIL               &   8.77 \\
WORDS & DOT                  &   8.67 \\
SUBJECTS & COMPARATIVE       &   8.39 \\
AUTHOR & D\_GOLDSMITH         &   8.04 \\
WORDS & ALH                  &   7.97 \\
WORDS & MANNED               &   7.97 \\
RELATED-TITLES & SETTLE      &   7.91 \\
RELATED-TITLES & CASE        &   7.91 \\
AUTHOR & R\_ZUBRIN            &   7.63 \\
AUTHOR & R\_WAGNER            &   7.63 \\
AUTHOR & H\_MORAVEC           &   7.63 \\
RELATED-AUTHORS & B\_DIGREGORIO &  7.63 \\
RELATED-AUTHORS & A\_RADFORD   &  7.63 \\
WORDS & LEE                   &  7.57 \\
WORDS & MORAVEC               &  7.57 \\
WORDS & WAGNER                &  7.57 \\
RELATED-TITLES & CONNECTIONIST  & 7.51 \\
RELATED-TITLES & BELOW        &  7.51 \\
\end{tabular}
\end{center}
\end{footnotesize}
\caption{\label{pos-profile}Sample Positive Profile Features}
\end{table}

\subsection{Producing, Explaining, and Revising Recommendations}

Once a profile is learned, it is used to predict the preferred ranking of the
remaining books based on posterior probability of a positive categorization,
and the top-scoring recommendations are presented to the user.

The system also has a limited ability to ``explain'' its recommendations by
listing the features that most contributed to its high rank.  For example,
given the profile illustrated above, {\sc Libra} presented the
explanation shown in Table~\ref{explain}.
\begin{table}[t]
\begin{center}
The Fabric of Reality: \\
The Science of Parallel Universes- And Its Implications \\
by David Deutsch recommended because:
\smallskip
\begin{footnotesize}
\begin{tabular}{llr}
Slot & Word & Strength \\
\hline
WORDS & MULTIVERSE  &            75.12 \\
WORDS & UNIVERSES   &            25.08 \\
WORDS & REALITY     &            22.96 \\
WORDS & UNIVERSE    &            15.55 \\
WORDS & QUANTUM     &            14.54 \\
WORDS & INTELLECT   &            13.86 \\
WORDS & OKAY        &            13.75 \\
WORDS & RESERVATIONS &           11.56 \\
WORDS & DENIES       &           11.56 \\
WORDS & EVOLUTION    &           11.02 \\
WORDS & WORLDS       &           10.10 \\
WORDS & SMOLIN       &           9.39 \\
WORDS & ONE          &           8.50 \\
WORDS & IDEAS        &           8.35 \\
WORDS & THEORY       &           8.28 \\
WORDS & IDEA         &           6.96 \\
SUBJECTS & REALITY   &           6.78 \\
TITLE & PARALLEL     &           6.76 \\
WORDS & IMPLY        &           6.47 \\
WORDS & GENIUSES      &          6.47 \\
\end{tabular}
\end{footnotesize} 
\end{center}
\caption{\label{explain}Sample Recommendation Explanation}
\end{table}
The strength of a cue in this case is multiplied by the number of times it
appears in the description in order to fully indicate its influence on the
ranking.  The positiveness of a feature can in turn be explained by listing the
user's training examples that most influenced its strength, as illustrated in
Table~\ref{explain-feature} where ``Count'' gives the number of times the
feature appeared in the description of the rated book.
\begin{table}[t]
\begin{center}
The word {\sc UNIVERSES} is positive due to your ratings:
\smallskip
\begin{footnotesize}
\begin{tabular}{lrr}
Title & Rating & Count \\
\hline
The Life of the Cosmos    &                        10   &   15 \\
Before the Beginning : Our Universe and Others &        8    &  7 \\
Unveiling the Edge of Time          &      10    &  3 \\
Black Holes : A Traveler's Guide    &    9    &  3 \\
The Inflationary Universe  &    9   &   2 \\
\end{tabular}
\end{footnotesize} 
\end{center}
\caption{\label{explain-feature}Sample Feature Explanation}
\end{table}

After reviewing the recommendations (and perhaps disrecommendations), the user
may assign their own rating to examples they believe to be incorrectly ranked
and retrain the system to produce improved recommendations.  As with {\it
relevance feedback} in information retrieval \cite{salton:jsis90}, this cycle
can be repeated several times in order to produce the best results.  Also, as
new examples are provided, the system can track any change in a user's
preferences and alter its recommendations based on the additional information.

\section{EXPERIMENTAL RESULTS}

\subsection{Methodology}

\subsubsection{Data Collection}

Several data sets were assembled to evaluate {\sc Libra}. The first two were
based on the first 3,061 {\it adequate-information} titles (books with at least
one abstract, review, or customer comment) returned for the subject search
``literature fiction.''  Two separate sets were randomly selected from this
dataset, one with 936 books and one with 935, and rated by two different
users. These sets will be called {\sc Lit1} and {\sc Lit2}, respectively.  The
remaining sets were based on all of the adequate-information Amazon titles for
``mystery'' (7,285 titles), ``science'' (6,177 titles), and ``science fiction''
(3,813 titles).  From each of these sets, 500 titles were chosen at random and
rated by a user (the same user rated both the science and science fiction
books).  These sets will be called {\sc Myst}, {\sc Sci}, and {\sc SF},
respectively.

In order to present a quantitative picture of performance on a realistic
sample; books to be rated where selected at random.  However, this means that
many books may not have been familiar to the user, in which case, the user was
asked to supply a rating based on reviewing the Amazon page describing the
book.  Table~\ref{data-info} presents some statistics about the data and
Table~\ref{ratings-info} presents the number of books in each rating
category.  Note that overall the data sets have quite different ratings
distributions.

\begin{table}
\begin{center}
\begin{tabular}{|l|ccc|}
\hline                                  
Data & Number Exs & Avg. Rating & \% Positive ($r>5$)\\
\hline
{\sc Lit1} & 936 & 4.19 & 36.3 \\
{\sc Lit2} & 935 & 4.53 & 41.2 \\
{\sc Myst} & 500 & 7.00 & 74.4 \\
{\sc Sci}  & 500 & 4.15 & 31.2 \\
{\sc SF}   & 500 & 3.83 & 20.0 \\
\hline
\end{tabular}
\end{center}
\caption{\label{data-info}Data Information}
\end{table}

\begin{table}
\begin{footnotesize}
\begin{center}
\begin{tabular}{|l|rrrrrrrrrr|}
\hline            
  & \multicolumn{10}{|c|}{Rating} \\                      
Data & 1 & 2 & 3 & 4 & 5 & 6 & 7 & 8 & 9 & 10 \\
\hline
{\sc Lit1} & 271 &  78 &  67 &  74 &  106 & 125 &  83 &  70 &  40 &  22 \\ 
{\sc Lit2} & 272 &  58 &  72 &  92 &  56 &  75 &  104 &  87 &  77 & 42 \\
{\sc Myst} & 73 &  11 &  7 &  8 &  29 &  46 &  45 &  64 &  66 &  151 \\
{\sc Sci}  & 88 &  94 &  62 &  49 &  51 &  53 &  35 &  31 &  16 &  21 \\
{\sc SF}   & 56 &  119 &  75 &  83 &  67 &  33 &  28 &  21 &  12 &  6 \\
\hline
\end{tabular}
\end{center}
\end{footnotesize}
\caption{\label{ratings-info}Data Rating Distributions}
\end{table}

\begin{table*}[t]
\begin{center}
\begin{tabular}{|lr|rrrrrrrrr|}
\hline                                  
Data       & N   & Acc  & Rec  & Pr   & Pr3  & Pr10 & F    & Rt3  & Rt10 & $r_s$ \\
\hline
{\sc Lit1} &   5 & 63.5 & 49.0 & 50.3 & 63.3 & 62.0 & 46.5 & 5.87 & 6.02 & 0.31 \\
{\sc Lit1} &  10 & 65.5 & 51.3 & 53.3 & 86.7 & 76.0 & 49.7 & 6.63 & 6.65 & 0.35 \\
{\sc Lit1} &  20 & 73.4 & 64.8 & 62.6 & 86.7 & 81.0 & 62.6 & 7.53 & 7.20 & 0.62 \\
{\sc Lit1} &  40 & 73.9 & 65.1 & 63.6 & 86.7 & 81.0 & 63.4 & 7.40 & 7.32 & 0.64 \\
{\sc Lit1} & 100 & 79.0 & 70.7 & 71.1 & 96.7 & 86.0 & 70.5 & 8.03 & 7.44 & 0.69 \\
{\sc Lit1} & 840 & 79.8 & 62.8 & 75.9 & 96.7 & 94.0 & 68.5 & 8.57 & 8.03 & 0.74 \\
\hline
{\sc Lit2} &   5 & 59.0 & 57.6 & 52.4 & 70.0 & 74.0 & 53.3 & 6.80 & 6.82 & 0.31 \\
{\sc Lit2} &  10 & 65.0 & 64.5 & 56.7 & 80.0 & 82.0 & 59.2 & 7.33 & 7.33 & 0.48 \\
{\sc Lit2} &  20 & 69.5 & 67.2 & 63.2 & 93.3 & 91.0 & 64.1 & 8.20 & 7.84 & 0.59 \\
{\sc Lit2} &  40 & 74.3 & 72.1 & 68.9 & 93.3 & 91.0 & 69.0 & 8.53 & 7.94 & 0.69 \\
{\sc Lit2} & 100 & 78.0 & 78.5 & 71.2 & 96.7 & 94.0 & 74.4 & 8.77 & 8.22 & 0.72 \\
{\sc Lit2} & 840 & 80.2 & 71.9 & 78.6 & 100.0 & 97.0 & 74.8 & 9.13 & 8.48 & 0.77 \\
\hline
{\sc Myst} &   5 & 73.2 & 83.4 & 82.1 & 86.7 & 89.0 & 81.5 & 8.20 & 8.40 & 0.36 \\
{\sc Myst} &  10 & 75.6 & 87.9 & 82.4 & 90.0 & 90.0 & 83.8 & 8.40 & 8.34 & 0.40 \\
{\sc Myst} &  20 & 81.6 & 89.3 & 86.4 & 96.7 & 91.0 & 87.3 & 8.23 & 8.43 & 0.46 \\
{\sc Myst} &  40 & 85.2 & 95.4 & 85.9 & 96.7 & 94.0 & 90.3 & 8.37 & 8.52 & 0.50 \\
{\sc Myst} & 100 & 86.6 & 95.2 & 87.2 & 93.3 & 94.0 & 90.9 & 8.70 & 8.69 & 0.55 \\
{\sc Myst} & 450 & 85.8 & 93.2 & 88.1 & 96.7 & 98.0 & 90.5 & 8.90 & 8.97 & 0.61 \\
\hline
{\sc Sci}  &   5 & 62.8 & 63.8 & 46.3 & 73.3 & 60.0 & 51.1 & 6.97 & 6.17 & 0.35 \\
{\sc Sci}  &  10 & 67.6 & 61.9 & 51.2 & 80.0 & 67.0 & 54.3 & 7.30 & 6.32 & 0.37 \\
{\sc Sci}  &  20 & 75.4 & 66.0 & 64.2 & 96.7 & 80.0 & 63.1 & 8.37 & 7.03 & 0.51 \\
{\sc Sci}  &  40 & 79.6 & 69.5 & 68.7 & 93.3 & 80.0 & 68.3 & 8.43 & 7.23 & 0.59 \\
{\sc Sci}  & 100 & 81.8 & 74.4 & 72.2 & 93.3 & 83.0 & 72.3 & 8.50 & 7.29 & 0.65 \\
{\sc Sci}  & 450 & 85.2 & 79.1 & 76.8 & 93.3 & 89.0 & 77.2 & 8.57 & 7.71 & 0.71 \\
\hline
{\sc SF}   &   5 & 67.0 & 38.3 & 32.9 & 40.0 & 29.0 & 28.2 & 5.23 & 4.34 & 0.02 \\
{\sc SF}   &  10 & 64.6 & 49.0 & 28.9 & 53.3 & 36.0 & 31.5 & 5.83 & 4.72 & 0.15 \\
{\sc SF}   &  20 & 71.8 & 45.8 & 37.4 & 66.7 & 37.0 & 37.8 & 6.23 & 5.04 & 0.21 \\
{\sc SF}   &  40 & 72.6 & 58.9 & 40.1 & 70.0 & 43.0 & 43.0 & 6.47 & 5.26 & 0.39 \\
{\sc SF}   & 100 & 76.4 & 65.7 & 46.2 & 80.0 & 56.0 & 52.4 & 7.00 & 5.75 & 0.40 \\
{\sc SF}   & 450 & 79.2 & 82.2 & 49.1 & 90.0 & 63.0 & 60.6 & 7.70 & 6.26 & 0.61 \\
\hline
 \end{tabular}
\end{center}
\caption{\label{results} Summary of Results}
\end{table*}

\subsubsection{Performance Evaluation}

To test the system, we performed 10-fold cross-validation, in which each data
set is randomly split into 10 equal-size segments and results are averaged over
10 trials, each time leaving a separate segment out for independent testing,
and training the system on the remaining data \cite{mitchell:book97}.  In order
to observe performance given varying amounts of training data, {\it learning
curves} were generated by testing the system after training on increasing
subsets of the overall training data.  A number of metrics were used to measure
performance on the novel test data, including:
\begin{itemize}
\item {\it Classification accuracy} (Acc): The percentage of examples
correctly classified as positive or negative.
\item {\it Recall} (Rec): The percentage of positive examples classified
as positive.
\item {\it Precision} (Pr): The percentage of examples classified as
positive which are positive.
\item {\it Precision at Top 3} (Pr3): The percentage of the 3 top ranked 
examples which are positive.
\item {\it Precision at Top 10} (Pr10): The percentage of the 10 top ranked 
examples which are positive.
\item {\it F-Measure} (F): A weighted average of precision and recall
frequently used in information retrieval:\\ $F = (2 \cdot Pr \cdot Rec) / (Pr + Rec)$
\item {\it Rating of Top 3} (Rt3): The average user rating assigned to the 3
top ranked examples.
\item {\it Rating of Top 10} (Rt10): The average user rating assigned to the 10
top ranked examples.
\item {\it Rank Correlation} ($r_s$): Spearman's rank
correlation coefficient between the system's ranking and that imposed by the
users ratings ($-1 \leq r_s \leq 1$); ties are handled using the method
recommended by \cite{finn:book96}.
\end{itemize}
The top 3 and top 10 metrics are given since many users will be primarily
interested in getting a few top-ranked recommendations.  Rank correlation gives
a good overall picture of how the system's continuous ranking of books agrees
with the user's, without requiring that the system actually predict the
numerical rating score assigned by the user.  A correlation coefficient of 0.3
to 0.6 is generally considered ``moderate'' and above 0.6 is considered
``strong.''

\subsection{Basic Results}

The results are summarized in Table~\ref{results}, where $N$ represents the
number of training examples utilized and results are shown for a number of
representative points along the learning curve.  Overall, the results are quite
encouraging even when the system is given relatively small training sets.  The
{\sc SF} data set is clearly the most difficult since there are very few
highly-rated books.

The ``top n'' metrics are perhaps the most relevant to many users.  Consider
precision at top 3, which is fairly consistently in the 90\% range after only
20 training examples (the exceptions are {\sc Lit1} until 70
examples\footnote{References to performance at 70 and 300 examples are based
on learning curve data not included in the summary in Table~\ref{results}.} and 
{\sc SF} until 450 examples).
Therefore, {\sc Libra}'s top recommendations are highly likely to be viewed
positively by the user.  Note that the ``\% Positive'' column in
Table~\ref{data-info} gives the probability that a randomly chosen example from
a given data set will be positively rated.  Therefore, for every data set, the
top 3 and top 10 recommendations are always substantially more likely than
random to be rated positively, even after only 5 training examples.

Considering the average rating of the top 3 recommendations, it is fairly
consistently above an 8 after only 20 training examples (the exceptions again are
{\sc Lit1} until 100 examples and {\sc SF}).  For every data set, the top 3 and
top 10 recommendations are always rated substantially higher than a randomly
selected example (cf. the average rating from Table~\ref{data-info}).

Looking at the rank correlation, except for {\sc SF}, there is at least a
moderate correlation ($r_s \geq 0.3$) after only 10 examples, and {\sc SF}
exhibits a moderate correlation after 40 examples.  This becomes a strong
correlation ($r_s \geq 0.6$) for {\sc Lit1} after only 20 examples, for {\sc
Lit2} after 40 examples, for {\sc Sci} after 70 examples, for {\sc Myst} after
300 examples, and for {\sc SF} after 450 examples.

\subsection{Results on the Role of Collaborative Content}

Since collaborative and content-based approaches to recommending have somewhat
complementary strengths and weaknesses, an interesting question that has
already attracted some initial attention \cite{balabanovic:cacm97,basu:aaai98}
is whether they can be combined to produce even better results.  Since {\sc
Libra} exploits content about related authors and titles that Amazon produces
using collaborative methods, an interesting question is whether this {\it
collaborative content} actually helps its performance.  To examine this issue,
we conducted an ``ablation'' study in which the slots for related authors and
related titles were removed from {\sc Libra}'s representation of book content.
The resulting system, called {\sc Libra-NR}, was compared to the original one
using the same 10-fold training and test sets.  The statistical significance
of any differences in performance between the two systems was evaluated
using a 1-tailed paired {\it t}-test requiring a significance level of
$p < 0.05$.

Overall, the results indicate that the use of collaborative content has a
significant positive effect.  
\begin{figure}[t]
     \epsfxsize=3.5in
     \epsfbox{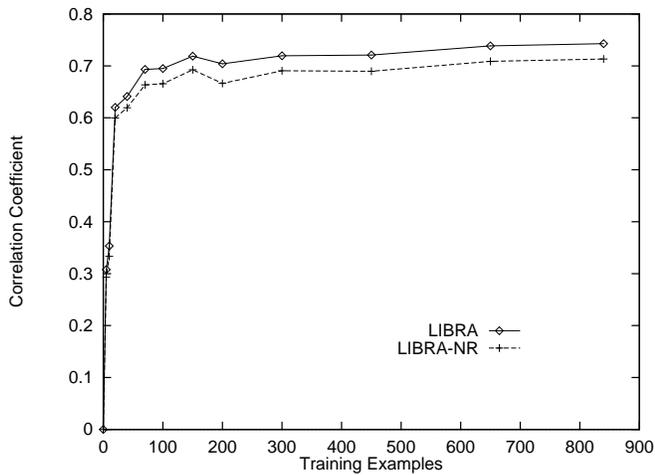}
\caption{\label{lit1-rs}{\sc Lit1} Rank Correlation}
\end{figure}
\begin{figure}[t]
     \epsfxsize=3.5in
     \epsfbox{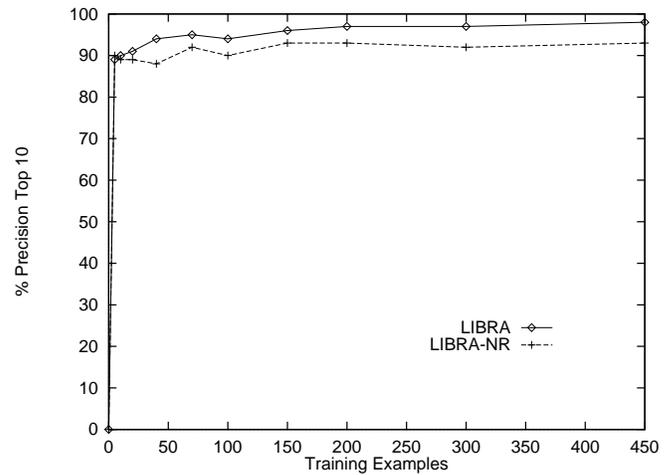}
\caption{\label{myst-pr10}{\bf {\sc Myst} Precision at Top 10}}
\end{figure}
\begin{figure}
     \epsfxsize=3.5in
     \epsfbox{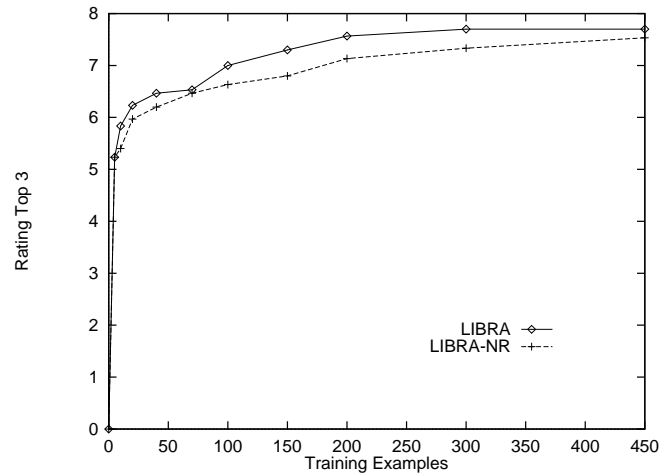}
\caption{\label{sf-rt3}{\bf {\sc SF} Average Rating of Top 3}}
\end{figure}
Figures~\ref{lit1-rs}, \ref{myst-pr10}, and \ref{sf-rt3}, show sample learning
curves for different important metrics for a few data sets. For the {\sc Lit1}
rank-correlation results shown in Figure~\ref{lit1-rs}, there is a consistent,
statistically-significant difference in performance from 20 examples onward.
For the {\sc Myst} results on precision at top 10 shown in
Figure~\ref{myst-pr10}, there is a consistent, statistically-significant
difference in performance from 40 examples onward.  For the {\sc SF} results on
average rating of the top 3, there is a statistically-significant difference at
10, 100, 150, 200, and 450 examples.  The results shown are some of the most
consistent differences for each of these metrics; however, all of the datasets
demonstrate some significant advantage of using collaborative content according
to one or more metrics.  Therefore, information obtained from collaborative
methods can be used to improve content-based recommending, even when the actual
user data underlying the collaborative method is unavailable due to privacy or
proprietary concerns.

\section{FUTURE WORK}

We are currently developing a web-based interface so that {\sc Libra} can be
experimentally evaluated in practical use with a larger body of users.  We plan
to conduct a study in which each user selects their own training examples,
obtains recommendations, and provides final informed ratings after reading one
or more selected books.

Another planned experiment is comparing {\sc Libra}'s content-based approach
to a standard collaborative method.  Given the constrained interfaces provided
by existing on-line recommenders, and the inaccessibility of the underlying
proprietary user data, conducting a controlled experiment using the exact same
training examples and book databases is difficult. However, users could be
allowed to use both systems and evaluate and compare their final
recommendations.\footnote{Amazon has already made significantly more income
from the first author based on recommendations provided by {\sc Libra} than
those provided by its own recommender system; however, this is hardly a
rigorous, unbiased comparison.}

Since many users are reluctant to rate large number of training examples,
various machine-learning techniques for maximizing the utility of small
training sets should be utilized.  One approach is to use unsupervised learning
over unrated book descriptions to improve supervised learning from a smaller
number of rated examples.  A successful method for doing this in text
categorization is presented in \cite{nigam:aaai98}.  Another approach is {\it
active learning}, in which examples are acquired incrementally and the system
attempts to use what it has already learned to limit training by selecting only
the most informative new examples for the user to rate \cite{cohn:mlj94}.
Specific techniques for applying this to text categorization have been
developed and shown to significantly reduce the quantity of labeled examples
required
\cite{lewis:ml94,liere:aaai97}.  

A slightly different approach is to advise users on easy and productive
strategies for selecting good training examples themselves.  We have found that
one effective approach is to first provide a small number of highly rated
examples (which are presumably easy for users to generate), running the system
to generate initial recommendations, reviewing the top recommendations for
obviously bad items, providing low ratings for these examples, and retraining
the system to obtain new recommendations.  We intend to conduct experiments on
the existing data sets evaluating such strategies for selecting training
examples.

Studying additional ways of combining content-based and collaborative
recommending is particularly important.  The use of collaborative content in
{\sc Libra} was found to be useful, and if significant data bases of both user
ratings and item content are available, both of these sources of information
could contribute to better recommendations
\cite{balabanovic:cacm97,basu:aaai98}.  One additional approach is to automatically add
the related books of each rated book as additional training examples with the
same (or similar) rating, thereby using collaborative information to expand the
training examples available for content-based recommending.

A list of additional topics for investigation include the following.
\begin{itemize}

\item Allowing a user to initially provide keywords that are of known interest (or
disinterest), and incorporating this information into learned profiles by
biasing the parameter estimates for these words \cite{pazzani:mlj97}.

\item Comparing different text-categorization algorithms: In addition to more
sophisticated Bayesian methods, neural-network and case-based methods could be
explored.

\item Combining content  extracted from multiple sources:
For example, combining information about a title from Amazon, BarnesAndNoble,
on-line library catalogs, etc.

\item Using full-text  as content: A digital library should be
able to efficiently utilize the complete on-line text, as well as abstracted
summaries and reviews, to recommend items.

\end{itemize}

\section{CONCLUSIONS}

The ability to recommend books and other information sources to users based on
their general interests rather than specific enquiries will be an important
service of digital libraries.  Unlike collaborative filtering, content-based
recommending holds the promise of being able to effectively recommend unrated
items and to provide quality recommendations to users with unique, individual
tastes.  {\sc Libra} is an initial content-based book recommender which uses a
simple Bayesian learning algorithm and information about books extracted from
the web to recommend titles based on training examples supplied by an
individual user.  Initial experiments indicate that this approach can
efficiently provide accurate recommendations in the absence of any information
about other users.

In many ways, collaborative and content-based approaches provide complementary
capabilities.  Collaborative methods are best at recommending reasonably
well-known items to users in a communities of similar tastes when sufficient
user data is available but effective content information is not.  Content-based
methods are best at recommending unpopular items to users with unique tastes
when sufficient other user data is unavailable but effective content
information is easy to obtain.  Consequently, as discussed above, methods for
integrating these approaches will perhaps provide the best of both worlds.

Finally, we believe that methods and ideas developed in machine learning
research \cite{mitchell:book97} are particularly useful for content-based
recommending, filtering, and categorization, as well as for integrating
with collaborative approaches \cite{billsus:ml98,basu:aaai98}.  Given
the future potential importance of such services to digital libraries, we look
forward to an increasing application of machine learning techniques to these
challenging problems.

\section{ACKNOWLEDGEMENTS}

Thanks to Paul Bennett for contributing ideas, software, and data, and to Tina
Bennett for contributing data.  This research was partially supported by the
National Science Foundation through grant IRI-9704943.

\bibliographystyle{plain}
\bibliography{dl99}

\end{document}